
\input phyzzx
\font\zfont = cmss10 scaled \magstep1
\def\ZZ{\hbox{\zfont Z\kern-.4emZ}}
\def\bigone{\hbox{1\kern -.23em {\rm l}}}
\def\CP{{\em CP }}
\def\CPT{{\em CPT }}
\def\SCIPP{Santa Cruz Institute for Particle Physics,\break
University of California, Santa Cruz, CA 95064}
\overfullrule 0pt
\Pubnum{SCIPP 92/16}
\date{May, 1992}
\pubtype{ T}     
\titlepage
\title{{
Of \CP and other Gauge Symmetries in String Theory}
\foot{Work supported in part by the U.S. Department of Energy.}}
\author{Michael Dine, Robert G. Leigh
and Douglas A. MacIntire}
\address{}
\SCIPP
\vskip 1.50cm
\vbox{
\centerline{\bf Abstract}
We argue that \CP is a gauge symmetry in string theory.
As a consequence, \CP cannot be explicitly broken either perturbatively
or non-pertubatively; there can be no non-perturbative \CP-violating
parameters.  String theory is thus an example of a theory
where all $\theta$ angles arise due to spontaneous \CP violation,
and are in principle calculable.
}

\submit{Physical Review Letters}
\endpage

\parskip 0pt
\parindent 25pt
\overfullrule=0pt
\baselineskip=18pt
\tolerance 3500

\endpage
\pagenumber=1

\REF\pq{R.D. Peccei and H.R. Quinn, Phys. Rev. Lett. {\bf 38},
1440 (1977); Phys. Rev. {\bf D16}, 1791 (1977).}
\REF\spontaneous{A. Nelson, Phys. Lett. {\bf 136B}, 387 (1984);
S.M. Barr, Phys. Rev. Lett. {\bf 53}, 329 (1984); Phys. Rev.
{\bf D30}, 1805 (1984); P.H. Frampton and T.W. Kephart,
Phys. Rev. Lett. {\bf 65}, 1549 (1990).}
There are two standard suggestions for solving the strong \CP
problem.  The most popular is the Peccei-Quinn symmetry,
and its associated axion.\refmark{\pq}
Another possibility, which has been
pursued by several authors, is to suppose that the underlying
laws of nature are \CP-conserving,
and \CP is spontaneously broken.\refmark{\spontaneous}
In particular, one assumes that the ``bare $\theta$" is zero; the
observable $\theta$ is then calculable.  The main difficulty
with this program is to understand why the observed $\theta$ is
in fact so small.  Usually one tries to arrange that, as a consequence
of (other) symmetries, $\theta$ vanishes at tree level, and that loop
corrections are suppressed by powers of small Yukawa couplings and the
like.

\REF\witten{E. Witten, Phys. Lett. {\bf B149}, 351 (1984).}
Witten noted some time ago that string theory possesses axions,
and that as a result it has the potential to solve the strong
\CP problem.\refmark{\witten}
Since then, there has been much discussion as to
whether this axion can remove all $\theta$ angles, whether
there exist other axions, whether the axions have suitable
decay constants, and whether the minimum of the axion
potential is necessarily at $\theta=0$.  But little or no attention
has been paid to the question of whether or not string theory might
in fact be a theory of the second kind, \ie one where the
underlying, microscopic theory preserves \CP, and the bare $\theta$
vanishes.

\REF\strominger{A. Strominger and E. Witten, Commun. Math.
Phys. {\bf 101} 341 (1985).}
\REF\gsw{M.B. Green, J. Schwarz and E. Witten, {\it Superstring
Theory}, Cambridge University Press, New York (1986).}
\REF\douglas{M. Douglas and S. Shenker, Nucl. Phys. {\bf B355},
635 (1990).}
It has been noted in the literature that in string perturbation
theory, \CP is a good symmetry, which can be spontaneously broken
by expectation values for various types of moduli and matter
fields.\refmark{\strominger,\gsw}
This does not answer the question, however, of whether or not the bare
$\theta$ vanishes.  For example, there has been much speculation
as to the possible existence of non-perturbative parameters
in string theory.\refmark{\douglas} A priori, if such parameters
exist (in the case of critical strings), some could be \CP-violating;
$\theta$ angles might then arise as functions of these parameters.
Indeed, $\theta$ angles
are in some sense the paradigms of non-perturbative parameters.
For example, in the case of the $E_8 \times E_8$ theory, compactified
to four dimensions, it is natural to ask whether one could obtain
two (or more) $\theta$ angles, only one of which could be removed
by the model-independent axion.

In this brief note, we argue that this cannot happen: if
string theory has non-perturbative parameters, they
are necessarily \CP-conserving.  String theory, as a result,
is a perfect example of a theory in which the bare $\theta$
vanishes as a consequence of symmetry.  The basis of this
argument is a very simple observation:  in string theory,
four-dimensional \CP transformations are {\it gauge transformations}.
As a result, provided simply that the theory {\it exists}, no
explicit breaking of the symmetry is possible, perturbatively
or non-perturbatively.  In the course of this discussion,
we will encounter some other amusing facts.  For example, we
will see that the $\ZZ_2$ symmetry of the $E_8 \times E_8$ theory
which interchanges the two $E_8$'s is itself a gauge symmetry,
and again is not susceptible to explicit breaking.

To understand in what sense \CP, in four-dimensional compactifications
of string theory, can be thought of as a gauge symmetry, consider
some features of the ten-dimensional
heterotic string theory.  This is a theory which violates
$P$ and conserves $C$.  In particular, the GSO condition, which
requires that spinors be (say) left-handed, violates parity.
It is perhaps helpful to understand this statement from a world-sheet
viewpoint.  The two-dimensional field theory which describes the
ten-dimensional theory has a symmetry under which one changes
the signs of the nine space-like coordinates $x^i$, and separately
those of their right-handed fermionic partners, $\psi^i$.  However,
the separate transformations do not commute with the BRST operator
(the physical state conditions), and thus cannot be symmetries
in space-time. Simultaneously changing the signs of both the
$x^i$'s and the $\psi^i$'s does respect the BRST symmetry, but
this condition does not respect GSO, which
involves a product of all the $\psi$'s.

Charge conjugation, on the other hand, is a good symmetry.
If we choose a Majorana basis for the Dirac matrices,
$C$ is just the instruction to take the complex conjugate of (space-time)
spinor fields.  The reality condition on these fields is obviously
invariant under this operation, as are the GSO and physical state
conditions.  In the left-moving sector, the effect of $C$ is most
easily understood in the bosonic formulation.  There it is just the
instruction to take $X^I \rightarrow -X^I$, $I=1, \dots , 16$.
This is obviously a symmetry of the world-sheet Lagrangian.  It has
the effect on the lattice of taking $p^I \rightarrow -p^I$.  Since
this is a symmetry of both the $O(32)$ and $E_8 \times E_8$ lattices,
it corresponds to a good symmetry in space-time.

It is very easy to see that $C$ is in fact a gauge transformation,
in either the $E_8 \times E_8$ or $O(32)$ theories.
Consider, for definiteness, the $O(32)$ case; the argument
is virtually identical for $E_8 \times E_8$.
Indeed the transformation $C$ can be viewed as a set of rotations in
sixteen 2-dimensional subspaces through an angle $\pi$, a transformation
that is obviously contained in $O(32)$. In the fermionic formulation,
if one works with $16$ complex $\lambda^i$'s, $X^I \rightarrow -X^I$
corresponds to $\lambda^i \rightarrow \lambda^{\bar i}$; this is just the
rotation we have described.

If the theory is compactified toroidally, $C$ must also reverse
the signs of {\it both} the left- and right-moving momenta associated
with the compact dimensions.  For compactification of an even
number of dimensions, however, this is obviously a proper
Lorentz transformation.

What about $P$ in lower dimensions?  It is well known that for toroidal
compactifications in string theory (and Kaluza-Klein theory), even
when one starts with a higher-dimensional theory which is $P$-violating,
the four-dimensional theory is $P$-conserving.  We don't expect that this
symmetry arises ``out of the air;" it must be one of the symmetries
of the original ten-dimensional theory -- indeed, it must be a
(proper) Lorentz transformation in that theory.  To see explicitly
what it is, it is helpful to consider a toroidal
compactification of the theory (without background
gauge or antisymmetric tensor fields), and to group the six internal
coordinates as three complex ones, $y^1 = x^4 + i x^5$, etc.
Then consider the transformation which reverses the signs of
$x^1,x^2,x^3$, $x^5$, $x^7$, and $x^9$ (and $\psi^1$, $\psi^2$, etc.).
This is a combination
of ordinary parity in four dimensions, times complex conjugation
of the $y^i$'s (and $\psi^i$'s); it commutes with GSO.
{}From a ten-dimensional perspective, it is a proper
Lorentz transformation.
It is easy to verify that on massless
fermions it has precisely the correct effect.
The reader who wishes to check this point may find it
convenient to adopt the following basis for the ten-dimensional
Dirac matrices
$$\Gamma^{\mu} = \gamma^{\mu} \otimes \bigone, \mu=0,\cdots, 3$$
$$\Gamma^I = \gamma^5 \otimes \gamma^I,I=1,\dots,6 .$$
Write the four dimensional $\gamma$ matrices in
a Weyl basis, and the $O(6)$ $\gamma$-matrices
in terms of creation and
annihilation operators (see \eg , Ref. [\gsw ]).
Then the $16$-component,
ten-dimensional spinors break up into pieces $u_{\alpha a}$,
$u^*_{\dot \alpha \bar a}$, where $\alpha$, $\dot \alpha$
denote left- and right-moving spinors, and $a$, $\bar a$
are indices referring to the $4$ and $\bar 4$ representations
of $O(6) \sim SU(4)$.  Grouping the $u_{\alpha a}$'s and
$u_{\dot \alpha \bar a}$'s into four four-component spinors, $\Psi^a$,
$P$ takes left-handed
fermions to their right-handed counterparts.  It is
straightforward to show, in addition,
that the corresponding vertex operators
are also suitably mapped into one another; similarly,
bosonic vertex operators have well-defined
transformation properties (\eg , scalars transform
as scalars or pseudoscalars, and gauge bosons transform
appropriately).

Of course, since both $C$ and $P$ are gauge transformations,
it follows that \CP is as well.  So far, however, we have only
illustrated these statements for toroidal compactifications.
Many other types of compactifications violate
$P$ and $C$ separately in four dimensions, while conserving
\CP.  A good example is provided by conventional Calabi-Yau
compactifications, with the so-called ``standard embedding of the
gauge group."\refmark{\gsw}
In these theories, while the $C$ and $P$ symmetries
which have been defined above are spontaneously broken
by the expectation values of the graviton and gauge fields,
the combination is conserved, for suitable values of the moduli.
Indeed, at the level of the $\sigma$-model which describes
such compactifications, $P \times C$ is precisely the CP
symmetry of Ref. [\strominger ].  The same construction also works for
symmetric orbifolds. Thus once again \CP can (almost certainly)
be thought of as a gauge symmetry.
While the complete space of four-dimensional string theories
is not known, and it is by no means clear that all such theories
can be obtained (for some limiting value of some moduli) by
solution of ten-dimensional field (or $\beta$-function) equations,
it is quite natural to suppose that this result is completely
general:  \CP is always a gauge symmetry in string theory.

\CP can, of course, be spontaneously broken in string theory, as
stressed long ago by Strominger and Witten.\refmark{\strominger}
\REF\ginsparg{P. Ginsparg, Phys. Rev. {\bf D35}, 648 (1987).}
Before speculating on how this might occur with sufficiently small
effective $\theta$, it is instructive to understand the absence
of multiple $\theta$ parameters in other, rather similar, ways.
Consider, for example, toroidal compactifications of the heterotic
string.  We have already remarked that one might worry that
there are different $\theta$'s for each low energy gauge
group, only one of which can be removed by the model-independent
axion. That this is not the case follows from our
discussion of \CP, but it can be seen another way. Consider first
the $E_8 \times E_8$ theory.  In this case, it is tempting to say
that one can add two $\theta$'s, one for each $E_8$.  But in
perturbation theory there is a $\ZZ_2$ which
relates the two $E_8$'s.  It is not hard to show that this is
a gauge symmetry, and thus the two $\theta$'s are necessarily
the same.  For example, in Ref. [\ginsparg\ ]
it was shown that by turning on a background expectation
value for certain gauge fields, one can map the $E_8 \times E_8$
theory continuously to the $O(32)$ theory.
But under this mapping, it is
a straightforward matter to check that the $\ZZ_2$ is mapped
into a particular $O(32)$ gauge transformation. That is, one
obtains the $\ZZ_2$ transformation by turning on background fields such
that one obtains the $O(32)$ theory, rotating the lattice by an $O(32)$
transformation, and then returning to the ($\ZZ_2$ transformed)
$E_8 \times E_8$ theory by again turning on certain background fields.

More generally, one can ask whether different $\theta$'s might appear
as one moves around in the moduli space, \eg , at points of enhanced
symmetry.  Again, the answer is no, as a consequence of gauge
invariance.  We are worried, here, about terms which do not
change as one moves around in the moduli space.  In particular, then,
we can ask about the coefficient of $F \tilde F$ for each
of the 22 $U(1)$ gauge bosons which exist everywhere in the moduli
space, associated with left-moving fields.
\foot{Note such $\theta$'s
are physically meaningful even at points where the $U(1)$'s are not
unified in non-Abelian groups.  To illustrate the point, consider
an $O(3)$ gauge theory with a scalar triplet.  If one gives the
scalar a small expectation value, the $O(3)$ is broken to $U(1)$,
but $\theta$-dependent effects are still present, albeit suppressed.}
There is a point in the moduli space where all $22$ of these symmetries
are unified in a single non-Abelian group.  At this point, gauge
invariance requires that all $\theta$'s be equal.

\REF\accident{G. Lazarides, C. Panagiotakopoulos and Q. Shafi,
Phys. Rev. Lett. {\bf 46},432 (1986);
J.A. Casas and G.G. Ross, Phys. Lett. {\bf B192}, 119 (1987).}
\REF\cinti{M. Dine, Talk at the Cincinnati Symposium in
Honor of the Retirement of Louis Witten, SCIPP preprint in
preparation.}
In view of these observations, one can envisage several scenarios
for solving the strong \CP problem in string theory. Our comments
here will be rather preliminary. The fact that \CP is spontaneously
broken and $\theta_{QCD}$ is in principle calculable does
not mean that it is small. For example, it could be that in string
theory there are several strong gauge groups, and that the effective
$\theta$'s for each of these groups is large. In this situation,
one must make sure that there are enough axions to cancel these
$\theta$'s. In addition to the model-independent axion, some further
approximate Peccei-Quinn symmetries must appear ``by accident,"
\eg , as a consequence of discrete symmetries.\refmark{\accident,
\cinti} This is the conventional view we referred to in the first
paragraph. Alternatively, perhaps the various $\theta$'s are simply
small with string theory realizing some version
of the ideas of Nelson and Barr\refmark\spontaneous\ ; the low
energy structure of string theory is sufficiently rich that this
might occur. One probably does not want \CP broken at too low
an energy; otherwise one cannot hope to inflate away the
associated domain walls. Implementing variants of these schemes
at high energies may require additional discrete symmetries;
fortunately these are common in string theory. These ideas will
be explored in a subsequent publication, where details of the
analyses reported here will also be presented.

The observation that \CP is a gauge symmetry raises an obvious
question, about which we will only make some timid speculations:
what about $T$ invariance, and \CPT?  We can not use precisely the
same sort of reasoning to argue that $T$ is a gauge symmetry as we
did for parity. The problem is that the would-be Lorentz transformation in
ten dimensions is not part of the proper Lorentz group.
However, from a stringy viewpoint,
the similarities between the four dimensional $T$
and $P$ are so striking that it is natural to speculate
that $T$, also, is a gauge tranformation.
Whether this would have
profound consequences, we do not presently know.
Unlike the case of field theory,
it is not easy to make a general statement about \CPT in string theory.
In perturbation theory, \CPT appears to hold, basically
because the space-time theory inherits \CPT from the world-sheet theory.
But it is not yet clear that \CPT need hold
non-perturbatively.  Of course, if $T$ as well as \CP is a gauge
transformation, this would ensure that \CPT is as well.

\bigskip
\noindent
{\bf Acknowledgements: }
We thank L. Dixon,
A. Nelson, P. Frampton, H. Haber and E. Witten for conversations,
and T. Banks and N. Seiberg for comments on the manuscript.
As we were completing this paper, we learned that
K. Choi, D. Kaplan and A. Nelson have noted that \CP is often a gauge
symmetry in Kaluza-Klein theories.
\refout
\end